\documentclass[aps,pre,twocolumn,superscriptaddress]{revtex4-2}

\usepackage[utf8]{inputenc}
\usepackage{graphicx}
\usepackage{float}
\usepackage{amsmath,amssymb}
\usepackage[colorlinks=true,linkcolor=blue,citecolor=blue,urlcolor=blue]{hyperref}
\usepackage{listings}

\begin{document}

\title{Methodology of signal spectral analysis in various Kron's model geometries}






\author{Artur K. Michalak}
\affiliation{Quantum Hardware Systems, 94-056 Lodz, Poland}
\affiliation{Practical Electronics and Microelectronics Scientific Student Association (PEMSSA), \\ Lodz University of Technology, \\ Department of Microelectronics and Computer Science (DMCS),  
 93-005 Lodz, Poland }

\author{Maciej Dolecki}
\affiliation{Quantum Hardware Systems, 94-056 Lodz, Poland}
\affiliation{Lodz University of Technology, Institute of Physics \\ 
\&  Practical Electronics and Microelectronics Scientific Student Association (PEMSSA), $   $ \\ Department of Microelectronics and Computer Science (DMCS) and Institute of Physics, 
93-005 Lodz, Poland
}


\author{Krzysztof Pomorski}
\affiliation{University of New Mexico, Albuquerque, NM 87-106, USA}
\affiliation{Quantum Hardware Systems, 94-056 Lodz, Poland}

\author{Wojciech Nowakowski}
\affiliation{Quantum Hardware Systems, 94-056 Lodz, Poland}
\affiliation{Lodz University of Technology, Institute of Physics,  Schroedinger Cat Association   \\
\&  Practical Electronics and Microelectronics Scientific Student Association (PEMSSA), $   $  \\ Department of Microelectronics and Computer Science (DMCS), 
93-005 Lodz, Poland}

\author{Eryk Halubek}
\affiliation{Lodz University of Technology, Institute of Physics,  Schroedinger Cat Association   \\
}
\affiliation{Quantum Hardware Systems, 94-056 Lodz, Poland}

\date{20-June-2025}

\begin{abstract}
This work presents experimental and theoretical comparison in the modelling of the quantum wave function of single-electron in semiconductor nanowire via classical analog electronics based hardware emulator with the use of Kron concept. 
Thus we are able to express the semiconductor single-electron devices of linear or closed circle topology as present in position-based qubits that
can be mapped essentially to one dimensional Kron model implemented experimentally.  
We have also represented two dimensional single-electron wave function in Krons model and point out future experiments to be conducted. 
\end{abstract}

\maketitle

\section{Introduction}
\begin{figure}
\begin{center}
    \includegraphics[width=1\linewidth]{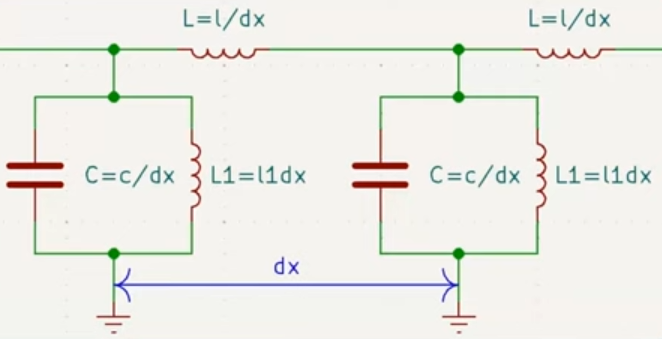}   
    \caption{Concept of Kron model \cite{PhysRev.67.39} encoding wave function dynamics by one dimensional electrical network with the phasor of voltages across capacitors being equivalent to wave function. }
\end{center}
\label{fig:Kron1}
\end{figure}
Currently there is a worldwide race in the development of quantum technologies and in particular in implementation of quantum computer.
Various technologies are considered, but the major candidates are usually associated with Josephson junction transmon based technology \cite{IBMQExp}
or semiconductor single-electron devices referring to semiconductor quantum dots, nanowires \cite{Fujisawa}, \cite{Spie} and cryogenic CMOS. All those scalable technologies are operational in deeply cryogenic environment what triggers the high experimental costs. 
Due to existence of various analog electronic based hardware emulators of quantum systems working at room temperature one can reduce those costs. 
The good example was delivered by G.Kron model \cite{PhysRev.67.39},\cite{condmat9040035} of the quantum system depicted in Fig.1,
in which the transmission line dynamics yield a system of two coupled linear ODE differential equations: one describing the relationship between the derivative of the voltage phasor at the given position ( at the given knot) and the current phasor at the corresponding posion (given knot) as by equation \ref{eq:1}, and by the next formula relating the derivative of the phasor current to the voltage as by equation \ref{eq:2}:

 \begin{eqnarray} \label{eq:1}
\frac{dV(x)}{dx} = \text{i}\omega L I(x),
\label{eq:2}
    \frac{dI(x)}{dx} = V (x)(-\text{i}\omega C+\frac{\text{i}}{\omega L^{}_{1}}).    
\end{eqnarray} Very last two formulas leads to the second order linear differential equation \ref{eq:3}
\begin{equation}\label{eq:3}
-\frac{1}{L}\frac{d^2\hat{V}(x)}{dx^2} +[\frac{1}{L^{}_{1}(x)}-\omega^2C ]\hat V (x)=0.
\end{equation} By assigning appropriate values to the individual components of the system \begin{math}
    \omega^2C^{}_{1} \times g= E,\space \frac{1}{L^{}_{1}(x)} \times g = V(x), \space \frac{1}{L} \times g= \frac{\hbar^2}{2m}
\end{math}, where \begin{math}
    g=1[\frac{J^2}{A^2}]
\end{math}, it is possible to derive the Schrödinger equation, where the voltage values at successive capacitor nodes correspond to a wave function scaled by a unit factor. For this reason, Kron's model can be effectively employed to simulate the wave function in more complex systems. However, there are cases where it is necessary to develop a model that simulates a simpler scenario, one that can be computed using a standard computational device. In this work we have presented numerical schemes to calculate voltage phasor at capactive elements in the case of classic linear Kron's model emulating position-based qubit in single electron regime [\hyperlink{2}{2}-\hyperlink{3}{3}], and in the case of closed semiconductor nanoring in single-electron regime and named as roton [\hyperlink{4}{4}]. Standard Toeplitz matrix representation of Schroedinger Hamiltonian \cite{Schroedinger} is used what immediately gives the corresponding energy eigenvalues and eigenfunctions that can be immediately translated into wavefunction dependence on position with time. 


This work presents a theoretical scheme for computation of signal spectra in one and in two dimensional Kron's model with specification of different computational methods.
This work also presents circular semiconductor nanowire in single electron regime defining so called roton that is experimentally implemented in Kron's model. Corresponding occupancy of eigenenergy levels is obtained for different frequency of signal applied to the system by external signal generator.  
 \begin{figure}
   \begin{center}
    \includegraphics[width=0.5\linewidth]{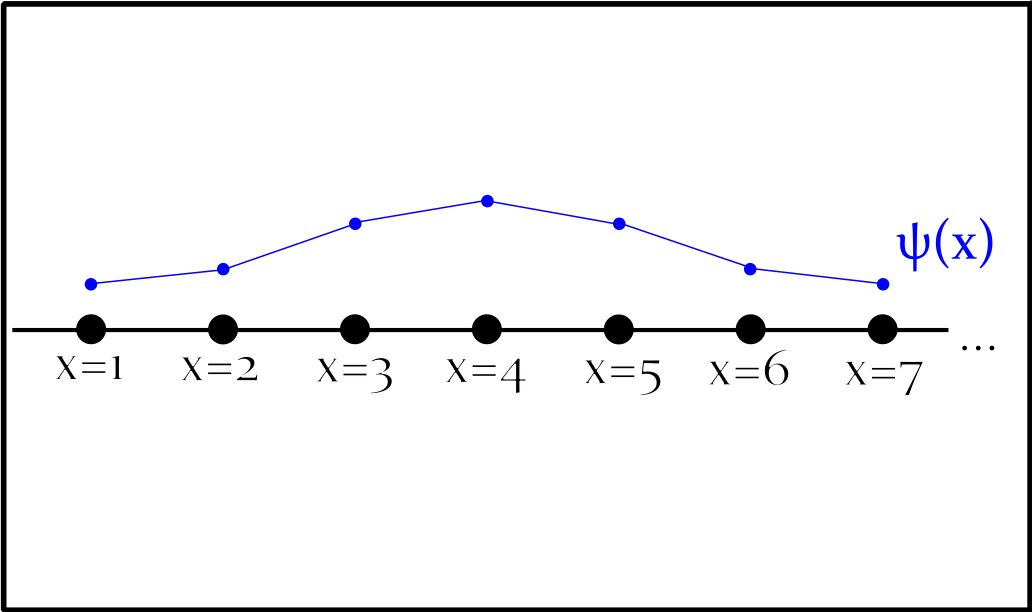}   
    \caption{One dimensional representation of position-based qubit \cite{Spie},\cite{Wannier} \emph{(linear Wannier qubit)} in single-electron regime with Kron RLC model \cite{condmat9040035} with 7 knots presented by black dots and sample of wave function obtained experimentally given by blue curve. }
    \end{center}
    \label{1s}
    \end{figure}
     \begin{figure}
     \begin{center}
      \includegraphics[width=0.5\linewidth]{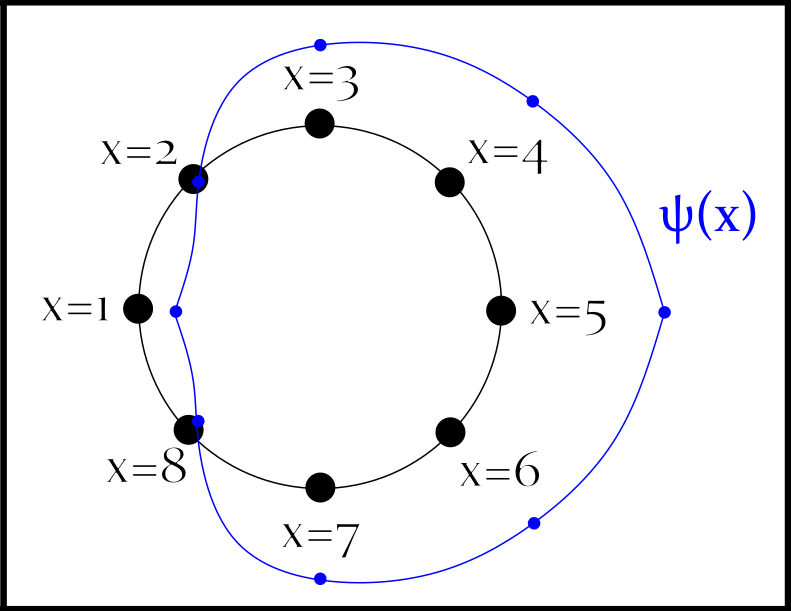}
    \caption{One dimensional representation of position-based circular qubit \cite{Spie},\cite{Wannier} \emph{(Roton Wannier qubit)} in single-electron regime with Kron RLC modeli with 8 knots presented by black dots and sample  of wave function obtained experimentally given by blue curve.}
     \end{center}
     \label{2s}
         \end{figure}
     \begin{figure}
   \begin{center}
    \includegraphics[width=0.6\linewidth]{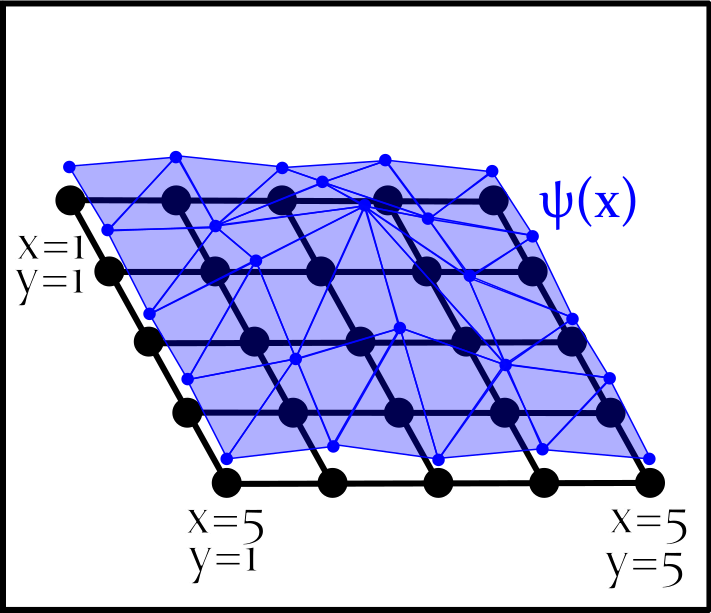}
    \caption{Experimental results for two dimensional Krons model with total number of 25 knots encoding lattice $N_x=5$ by $N_y=5$ presented by black dots (knots) and experimentally obtained wave function given by blue curve.}
    \end{center}
    \label{3s}
   \end{figure}

\section{Method 1 and 2: Two computational schemes in calculation of the voltage phasor}
Two methods can be identified for the numerical calculation of the voltage at successive nodes. Method 1 involves use of coupled equations
1 
and 2, 
 while the Method 2 relies on the final solution of equation 3
 via Toeplitz matrix use.

The first method 1 requires approximating the derivatives in the system of equations using finite differences. Formulas (4) and (5) are transformation of the two ODE's (1) and (2). In this case: 
\begin{equation} \label{eq:4}
     \frac{V(s+1)-V(s)}{\Delta x}= \text{i}\omega L I(s) 
\end{equation}
\begin{equation} \label{eq:5}
    \frac{I(s+1)-I(s)}{\Delta x} = (-\text{i}\omega C+\frac{\text{i}}{\omega L^{}_{1}})V(s)
\end{equation}
From this, successive values of voltage [eq. (6)] and current [eq. (7)] at the nodes can be calculated: 
\begin{equation} \label{eq:6}
    V(s+1)=V(s)+ \text{i}\omega L \Delta x I (s),  
\end{equation}
\begin{equation} \label{eq:7}
    I(s+1) = I(s)+V (s)(-\text{i}\omega C+\frac{\text{i}}{\omega L_1(s)^{}_{1}})\Delta x
\end{equation}
These equations can be solved by defining the values of V(1) and I(1), as well as the passive elements of the system. This method allows for the numerical approximation of the functions V(x) and I(x). Both functions are, of course, complex. 
The second method approximates the second derivative in equation (3) for creating the formula (8). Then: \begin{equation}
    -\frac{1}{L}\frac{ V(s+2) -2V(s+1) + V(s)}{\Delta x^2} +[-\omega^2C +\frac{1}{L^{}_{1}(s)}]V (s)=0
\end{equation}
From where we can calculate the the following values of the Voltage function. This is shown in the equation number (9):\begin{equation}
    V(s+2)= 2V(s+1) +[-L\omega^2C\Delta x^2 +\frac{L\Delta x^2}{L^{}_{1}(s)}-1]V (s)
\end{equation}
We can use equation (4) to generate the value of Voltage on the second node whith using a initial values in the system [eq. (10)].: \begin{equation}
V(2)=V(1)+ I (1)i\omega L \Delta x
\end{equation}

This method is practically identical to the first. The only difference is that it calculates  successive values of V(x) by neglecting the often insignificant function I(x).
%


\section{Method 3: The Hamiltonian matrix eigenvalue method of calculating the voltage phasor in linear Wannier qubit in Kron model}
By transforming equation (1), a general expression for the Hamiltonian of the system can be obtained in the (11)th equation: 
\begin{equation}
    \hat{H}=[-\frac{1}{L}\frac{d^2}{dx^2}     +\frac{1}{L^{}_{1}(x)}]
\end{equation}

Using finite difference methods, the second derivative in this equation can be approximated by a square matrix of size N×N, where N is the number of nodes in the constructed Kron model. The function $L_1(x)$ can also be represented as a matrix of the same dimensions. After substituting the values, the resulting expression takes the following form: 
\begin{equation}
    \hat{H}=-\frac{1}{L}\frac{1}{\Delta x^2} 
    \begin{bmatrix}
    -2&1&0&0&...&0 \\
    1&-2&1&0&...&0 \\
    0&1&-2&1&...&0 \\
     0&0&1&-2&...&0 \\
    ...&...&...&...&...&... \\
    0&0&0&0&...&-2 \\
\end{bmatrix}    
+\frac{1}{L^{}_{1}(x)}
\end{equation}
\begin{equation}
\frac{1}{L^{}_{1}(x)}=
\begin{bmatrix}
        \frac{1}{L^{}_{1^{}_{1}}}&0&0&0&...&0 \\
         0&\frac{1}{L^{}_{1^{}_{2}}}&0&0&...&0 \\
         0&0&\frac{1}{L^{}_{1^{}_{3}}}&0&...&0 \\
         0&0&0&\frac{1}{L^{}_{1^{}_{4}}}&...&0 \\
         ...&...&...&...&...&... \\
    0&0&0&0&...&\frac{1}{L^{}_{1^{}_{N}}} \\
\end{bmatrix}.
\end{equation}

Both the formulas (12) and (13) are defying full form of the Hamiltonian.
If we defined matrices $V_q$ and $D_q$, where Vq is the matrix of eigenvectors of the Hamiltonian and $D_q$ is the matrix of its eigenvalues, the following relations (14) and (15) holds: 
\begin{equation}
    D^{}_{q}=\begin{bmatrix}
        E^{}_{1}&0&0&0&...&0 \\
    0&E^{}_{2}&0&0&...&0 \\
    0&0&E^{}_{3}&0&...&0 \\
    0&0&0&E^{}_{4}&...&0 \\
    ...&...&...&...&...&... \\
    0&0&0&0&...&E^{}_{N} \\
    \end{bmatrix}
\end{equation}
\begin{equation}
    V^{}_{q}=\begin{bmatrix}
    \psi^{}_{1^{}_{1}}&\psi^{}_{1^{}_{2}}&\psi^{}_{1^{}_{3}}&\psi^{}_{1^{}_{4}}&...&\psi^{}_{1^{}_{K}} \\
    \psi^{}_{2^{}_{1}}&\psi^{}_{2^{}_{2}}&\psi^{}_{2^{}_{3}}&\psi^{}_{2^{}_{4}}&...&\psi^{}_{2^{}_{K}} \\\psi^{}_{3^{}_{1}}&\psi^{}_{3^{}_{2}}&\psi^{}_{3^{}_{3}}&\psi^{}_{3^{}_{4}}&...&\psi^{}_{3^{}_{K}} \\\psi^{}_{3^{}_{1}}&\psi^{}_{3^{}_{2}}&\psi^{}_{3^{}_{3}}&\psi^{}_{3^{}_{4}}&...&\psi^{}_{3^{}_{K}} \\
    ...&...&...&...&...&... \\
    \psi^{}_{N^{}_{1}}&\psi^{}_{N^{}_{2}}&\psi^{}_{N^{}_{3}}&\psi^{}_{N^{}_{4}}&...&\psi^{}_{N^{}_{K}} \\
    \end{bmatrix}
\end{equation}

The last equation can be also written as (16):
\begin{equation}
    V^{}_{q}=\begin{bmatrix}
    \Psi^{}_{E1}\\
    \Psi^{}_{E2}\\
    \Psi^{}_{E3}\\
    \Psi^{}_{E4}\\
    ...\\
    \Psi^{}_{EN}\\
    \end{bmatrix}
\end{equation}

The formula (14) shows us the energy levels of the system (En), which are fulfilling equation (17).
\begin{equation}
    E = p_{E1}E_{1} +p_{E2} E_{2} + ... +p_{En} E_{n} ,
\end{equation}
where 
\begin{math}
    p^{}_{En}
\end{math}
is possibility of occurring the En state in the system.
We can also write:
\begin{eqnarray}
 \psi(x) =  \nonumber \\
 e^{i\gamma^{}_{1}}\sqrt{p^{}_{E1}} \psi ^{}_{E1}(x) +e^{i\gamma^{}_{2}}\sqrt{p^{}_{E2}} \psi ^{}_{E2}(x)...+e^{i\gamma^{}_{n}}\sqrt{p^{}_{En}}  \psi ^{}_{En}(x) \nonumber \\    
\end{eqnarray}
and:
\begin{eqnarray}
    e^{i\gamma^{}_{n}}\sqrt{p^{}_{En}}= 
     = \int_{-\infty}^{+\infty} (\psi^{T}_{En}(x))^* \psi(x)\ dx .
\end{eqnarray}
Equation (18) defines wave functions of energy levels and shows the wave function of the system as the superposition of the following wave functions of energies. The (19) formula gives connection of the constants in eq. (18) and the matrices of the wave functions. 
In our finite scheme we can write the last formula as the equation 
\begin{equation}
    \int_{-\infty}^{+\infty} (\psi^{T}_{En}(x))^* \psi(x)\ dx = \sum_{\substack{k=1 \\}}^{K} \psi^{*}_{En}(x_k)\psi(x_k) \Delta x =  e^{i\gamma^{}_{n}}\sqrt{p^{}_{En}}, 
\end{equation}
%
%
where 
\begin{math}
    \Delta x
\end{math} 
is a distance between 2 nodes and can be calculated from the (3) formula and the Schrodringer's equation.
After a few transformations of the (16), (18) and (20) formulas we can calculate the wave function:
\begin{equation}
    \psi(n)=\begin{bmatrix}
   \frac{\displaystyle\sum_{\substack{x=1 \\ x \ne 1}}^{N} (\displaystyle\sum_{\substack{k=1 \\ }}^{K} \psi^{*}_{Ek}(x)\psi_{Ek}(1))\psi(x)}{\displaystyle\frac{1}{\Delta x}-\displaystyle\sum_{\substack{k=1 \\}}^{K}|\psi_{Ek}(1)|^2 } \\ \\\frac{\displaystyle\sum_{\substack{x=1 \\ x \ne 2}}^{N} (\displaystyle\sum_{\substack{k=1 \\ }}^{K} \psi^{*}_{Ek}(x)\psi_{Ek}(2))\psi(x)}{\displaystyle\frac{1}{\Delta x}-\displaystyle\sum_{\substack{k=1 \\ }}^{K}|\psi_{Ek}(2)|^2 } \\ \\... \\ \\\frac{\displaystyle\sum_{\substack{x=1 \\ x \ne N}}^{N} (\displaystyle\sum_{\substack{k=1 \\ }}^{K} \psi^{*}_{Ek}(x)\psi_{Ek}(N))\psi(x)}{\displaystyle\frac{1}{\Delta x}-\displaystyle\sum_{\substack{k=1 \\ }}^{K}|\psi_{Ek}(N)|^2 } \\
    \end{bmatrix}
\end{equation} \[\]
Matrix (21) can be numerical solved. Equation (20) can be written as:
\begin{equation}
    p_{En} = |\sum_{\substack{k=1 \\}}^{K} \psi^{*}_{En}(x_k)\psi(x_k) \Delta x|^2
\end{equation}
%
%
%
%
\section{Method 4: The Hamiltonian matrix eigenvalue method of calculating the voltage phasor in roton qubit}
The only difference between way of calculating this method in line form of space and in roton is in the Hamiltonian. We can write it in changed form in the (23) formula:
\begin{equation}
\hat{H}=-\frac{1}{L}\frac{1}{\Delta x^2} 
\begin{bmatrix}
    -2&1&0&0&...&\mathbf{1}\\
    1&-2&1&0&...&0 \\
    0&1&-2&1&...&0 \\
     0&0&1&-2&...&0 \\
    ...&...&...&...&...&... \\
    \mathbf{1}&0&0&0&...&-2 \\
\end{bmatrix}    +\frac{1}{L^{}_{1}(x)} .
\end{equation} 



%
\section{Method 5: The Hamiltonian matrix eigenvalue method of calculating the voltage phasor two dimensional Krons model}
\twocolumngrid
It is possible to define a wave function that depends on more than one position index. In this study, we consider a system with two position indices (coordinates), 

\begin{math}
x
\end{math} 
and 
\begin{math}
y\end{math}, comprising\begin{math}
X \times Y\end{math} nodes. 
\begin{equation}
    -\frac{\hbar^2}{2m}\Delta\psi_{(x,y)}+\hat{V}_{(x,y)}\psi_{(x,y)} = E\psi_{(x,y)},
\end{equation}
where 
\begin{math}
    \Delta
\end{math} in formula (24) is a Laplace operator, which meets the condition:
\begin{equation}
    \Delta = [\frac{\delta^2}{\delta x^2}+\frac{\delta^2}{\delta y^2}] \equiv [\hat{\frac{\delta^2}{\delta x^2}} \otimes \hat{I_A}+\hat{I_B}  \otimes \hat{\frac{\delta^2}{\delta y^2}}] ,
\end{equation}
where the 
\begin{math}
    \otimes
\end{math} stands for kronecker's product. In eq. (25) we're defying:
\begin{equation}
    \hat{\frac{\delta^2}{\delta x^2}}=\frac{1}{\Delta x^2} \begin{bmatrix}
        -2 &1&0&0&...&0\\1 &-2&1&0&...&0\\0 &1&-2&1&...&0\\0 &0&1&-2&...&0\\... &...&...&...&...&...\\0 &0&0&0&...&-2
    \end{bmatrix}
\end{equation}
\begin{equation}
    \hat{\frac{\delta^2}{\delta y^2}}=\frac{1}{\Delta y^2} \begin{bmatrix}
        -2 &1&0&0&...&0\\1 &-2&1&0&...&0\\0 &1&-2&1&...&0\\0 &0&1&-2&...&0\\... &...&...&...&...&...\\0 &0&0&0&...&-2
    \end{bmatrix}
\end{equation}
\begin{equation}
     \hat{I_A}=\begin{bmatrix}
        1 &0&0&0&...&0\\0 &1&0&0&...&0\\0 &0&1&0&...&0\\0 &0&0&1&...&0\\... &...&...&...&...&...\\0 &0&0&0&...&1
    \end{bmatrix}
\end{equation}
\begin{equation}
     \hat{I_B}=\begin{bmatrix}
        1 &0&0&0&...&0\\0 &1&0&0&...&0\\0 &0&1&0&...&0\\0 &0&0&1&...&0\\... &...&...&...&...&...\\0 &0&0&0&...&1
    \end{bmatrix}.
\end{equation}
Matrices in (26) and (29) equations have same size 
(X\begin{math}
    \times
\end{math}X) so as matrices in (27) and (28) eq. (Y\begin{math}
    \times
\end{math}Y). 
We can define two dimensional Potential in the (30) formula:
\begin{eqnarray}
    \hat{V}_{(x,y)} \equiv \hat{V}_{(x)} \otimes \hat{V}_{(y)} = \nonumber \\
 =
    \begin{bmatrix}
        V_{(x_1,y_1)} &0&0&...&0\\0 &V_{(x_1,y_2)}&0&...&0\\0 &0&V_{(x_1,y_3)}&...&0\\... &...&...&...&...\\0 &0&0&...&V_{(x_X,y_Y)}
    \end{bmatrix}.
\end{eqnarray}
%
%
By applying the transformations introduced in the Introduction and using equations (24)–(30), the Hamiltonian of the two-dimensional system can be expressed as follows:
%
\begin{eqnarray}
    \small\hat{H}=-\frac{1}{L}(\frac{1}{\Delta x^2}+\frac{1}{\Delta y^2})
    \begin{bmatrix}
         -2&1&0&...&0\\1 &-2&1&...&0\\0 &1&-2&...&0\\... &...&...&...&...\\0 &0&0&...&-2
    \end{bmatrix}+ \nonumber \\
    \begin{bmatrix}
        \frac{1}{L_{1(x_1,y_1)}} &0&0&...&0\\0 &\frac{1}{L_{1(x_1,y_2)}}&0&...&0\\0 &0&\frac{1}{L_{1(x_1,y_3)}}&...&0\\... &...&...&...&...\\0 &0&0&...&\frac{1}{L_{1(x_X,y_Y)}}
    \end{bmatrix}
\end{eqnarray} .
%
%
If we define matrices \begin{math} V_q \end{math} and \begin{math} D_q \end{math}, where \begin{math} V_q \end{math} is the matrix of eigenvectors of the Hamiltonian and \begin{math} D_q \end{math} is the matrix of its eigenvalues, the following relations (32) and (33) hold: 
\begin{equation}
    V^{}_{q}=
    \begin{bmatrix}
    \Psi^{}_{E1(x,y)}\\
    \Psi^{}_{E2(x,y)}\\
    \Psi^{}_{E3(x,y)}\\
    \Psi^{}_{E4(x,y)}\\
    ...\\
    \Psi^{}_{EN(x,y)}\\
    \end{bmatrix}.
\end{equation} 
We can write it also as:
\begin{equation}
    V^{}_{q}=\begin{bmatrix}
    \Psi^{}_{E1(x)} \otimes \Psi^{}_{E1(y)}\\
    \Psi^{}_{E2(x)} \otimes \Psi^{}_{E2(y)}\\
    \Psi^{}_{E3(x)} \otimes \Psi^{}_{E3(y)}\\
    ...\\
    \Psi^{}_{EXY(x)} \otimes \Psi^{}_{EXY(y)}\\
    \end{bmatrix}.
\end{equation}
The only feasible way to separate the
\begin{math}
    \psi_{E(x,y)}
\end{math}  
vectors into two one-dimensional functions is through numerical computation.
Then we can use next steps from the second section separately for both 
\begin{math}
    \psi_{E(x)}
\end{math} and 
\begin{math}
    \psi_{E(y)}
\end{math} functions. The solutions were writed in (35) and (36) equations.
\begin{equation}
    \psi(x)=\begin{bmatrix}
   \frac{\displaystyle\sum_{\substack{x=1 \\ x \ne 1}}^{X} (\displaystyle\sum_{\substack{k=1 \\ }}^{K} \psi^{*}_{Ek}(x)\psi_{Ek}(1))\psi(x)}{\displaystyle\frac{1}{\Delta x}-\displaystyle\sum_{\substack{k=1 \\}}^{K}|\psi_{Ek}(1)|^2 } \\ \\\frac{\displaystyle\sum_{\substack{x=1 \\ x \ne 2}}^{X} (\displaystyle\sum_{\substack{k=1 \\ }}^{K} \psi^{*}_{Ek}(x)\psi_{Ek}(2))\psi(x)}{\displaystyle\frac{1}{\Delta x}-\displaystyle\sum_{\substack{k=1 \\ }}^{K}|\psi_{Ek}(2)|^2 } \\ \\... \\ \\\frac{\displaystyle\sum_{\substack{x=1 \\ x \ne X}}^{X} (\displaystyle\sum_{\substack{k=1 \\ }}^{K} \psi^{*}_{Ek}(x)\psi_{Ek}(X))\psi(x)}{\displaystyle\frac{1}{\Delta x}-\displaystyle\sum_{\substack{k=1 \\ }}^{K}|\psi_{Ek}(X)|^2 } \\
    \end{bmatrix}
\end{equation}
\begin{equation}
    \psi(y)=
    \begin{bmatrix}
   \frac{\displaystyle\sum_{\substack{y=1 \\ y \ne 1}}^{Y} (\displaystyle\sum_{\substack{k=1 \\ }}^{K} \psi^{*}_{Ek}(y)\psi_{Ek}(1))\psi(y)}{\displaystyle\frac{1}{\Delta y}-\displaystyle\sum_{\substack{k=1 \\}}^{K}|\psi_{Ek}(1)|^2 } \\ \\\frac{\displaystyle\sum_{\substack{y=1 \\ y \ne 2}}^{Y} (\displaystyle\sum_{\substack{k=1 \\ }}^{K} \psi^{*}_{Ek}(y)\psi_{Ek}(2))\psi(y)}{\displaystyle\frac{1}{\Delta y}-\displaystyle\sum_{\substack{k=1 \\ }}^{K}|\psi_{Ek}(2)|^2 } \\ \\... \\ \\\frac{\displaystyle\sum_{\substack{y=1 \\ y \ne Y}}^{Y} (\displaystyle\sum_{\substack{k=1 \\ }}^{K} \psi^{*}_{Ek}(y)\psi_{Ek}(Y))\psi(y)}{\displaystyle\frac{1}{\Delta y}-\displaystyle\sum_{\substack{k=1 \\ }}^{K}|\psi_{Ek}(Y)|^2 } \\
    \end{bmatrix}
\end{equation}
We can define the final two dimensional wave function as:
\begin{equation}
    \psi_{(x,y)} \equiv \psi_{(x)} \otimes \psi_{(y)}.
\end{equation}
%
%
%

\begin{figure}
\begin{center}
    \includegraphics[width=1\linewidth]{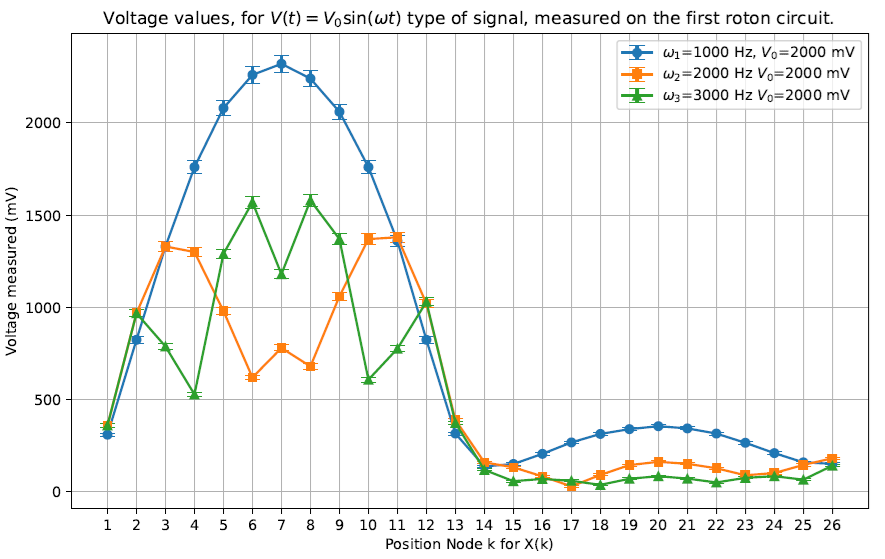}
    \caption{Various occupancy of energy levels as function of signal generator for semiconductor roton qubit in single-electron regime implemented in Kron's model, where horizontal axis corresponds to k-th given node and vertical axis corresponds to voltage phasor amplitude $|\hat{V}(k)|$. }
\end{center}
\label{fig:roton1}
\end{figure}

\begin{figure}
\begin{center}
    \includegraphics[width=1\linewidth]{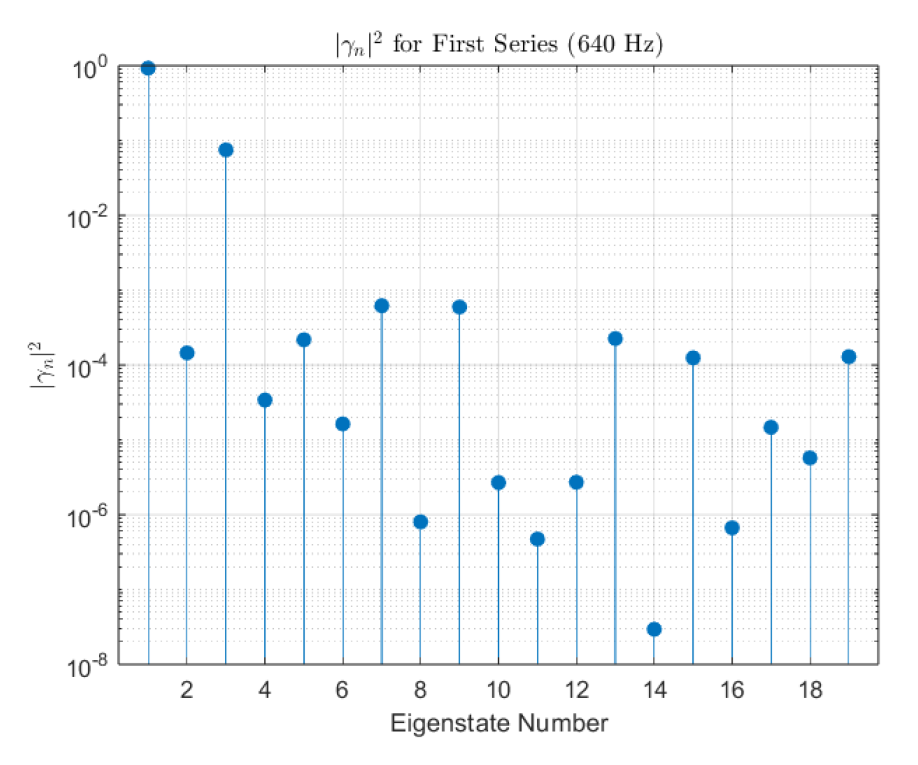}
        \caption{Spectral analysis of semiconductor roton qubit implemented in Kron's model from Fig.5. pointing the occupancy of the ground state  what occurs at frequencies of 640Hz. } 
\end{center}
\label{fig:roton2}
\end{figure}

\begin{figure}
\begin{center}
    \includegraphics[width=1\linewidth]{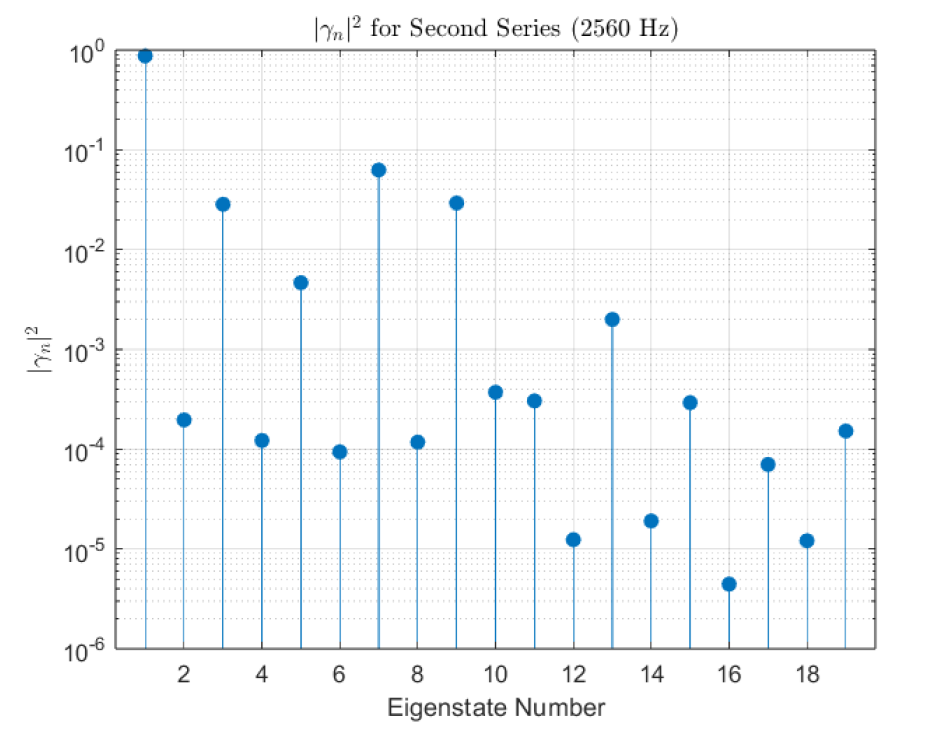}
        \caption{Spectral analysis of semiconductor roton qubit implemented in Kron's model from Fig.5. pointing the occupancy of the ground state and with certain presence of the excited states what occurs at frequencies of 2560Hz. } 
\end{center}
\label{fig:roton3}
\end{figure}

\begin{figure}
\begin{center}
    \includegraphics[width=1\linewidth]{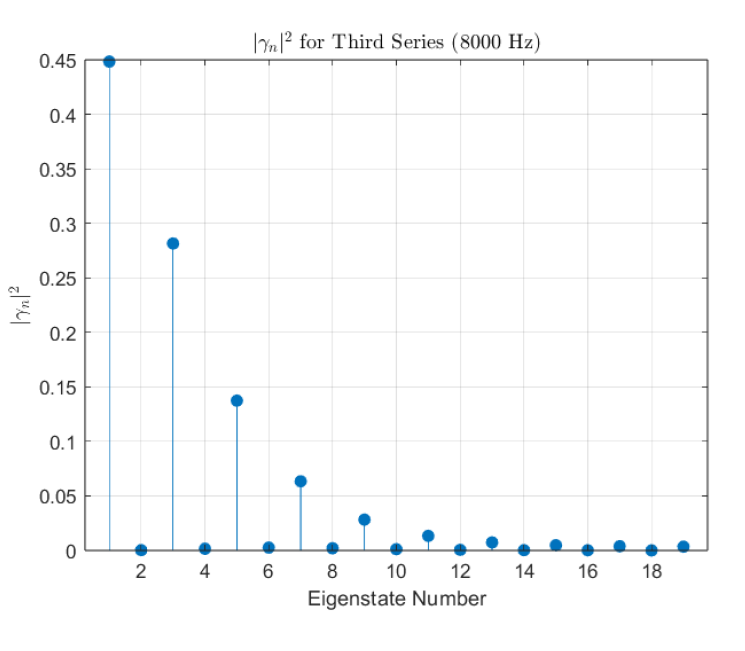}
        \caption{Spectral analysis of semiconductor roton qubit implemented in Kron's model from Fig.5. pointing the occupancy of the thermalized ground state corresponding to the Boltzman distribution, but only for odd k-th eigenergies what occurs at frequencies 8000Hz.} 
\end{center}
\label{fig:roton4}
\end{figure}
\section{Experiment with semiconductor roton in Kron's methodology}
The Kron model describing semiconductor nanowire in single-electron regime with 24 nodes was used basing on the circuit depicted in Fig.1, but with additional periodic boundary conditions. Two potential rectangular wells were defined and single external signal generator was used and connected to the node 7 that corresponds to the maxima of blue curve depicted in Fig.5. One could spot the that voltage phasor across capacitances were localized in two potential minima that corresponds to the Schroedinger wave-function solutions. Presence of ground state and excited energy states were identified as function of frequency of signal generator what is depicted in Fig.6-8.  The Boltzmann distribution of occupancy of odd energy levels was detected as specified by Fig.8. 
Therefore basic features of semiconductor single-electron qubit in circular geometry are confirmed in Kron's classical analog circuit representation at room temperature as proved by conducted experiments.  

\section{Discussion and outlook}
The presented experimental results validates the Kron's model effective representation of semiconductor nanowire in single-electron regime constituting semiconductor roton. The usage of few signal generators with time-dependent signal pattern shall validate Kron's model in all circumstances of qubit operation. The first author handled the theoretical derivations, developed the code for generating results based on the implemented methods, analyzed the data, and built the experimental system. The second author processed the measurement data and contributed to data analysis. The third author was responsible for the conceptual framework and theoretical basis of the study. The fourth author worked on both the theoretical and experimental aspects.
The last author have generated and processed the experimental data presented by Fig.5-8. 
\section{Acknowledgment}
We would like to thank to dr Mariusz Jankowski and to Practical Electronics and Microelectronics Scientific Student Association (PEMSSA) from Department of Microelectronics and Computer Science (DMCS) at Lodz University of Technology for various consultations with Authors and help in various experimental stages and critical comments. 
\onecolumngrid
\hypertarget{7}{}
\section{references}

\end{document}